\documentclass[a4paper,12pt]{article}

\usepackage[left=2.5cm,right=2.5cm,top=2.5cm,bottom=2.5cm]{geometry}
\usepackage{graphicx}
\usepackage{amssymb}
\usepackage{multirow}
\usepackage{amsmath}
\usepackage{amsthm}
\usepackage{psfrag}
\usepackage{setspace}
\usepackage{subcaption}
\usepackage[colorlinks=true,bookmarks=true]{hyperref}
\usepackage{breakurl} 
\usepackage{float}
\usepackage{cite}
\usepackage{authblk}
\usepackage{color}

\usepackage{mathtools}

\DeclarePairedDelimiterX\braket[2]{\langle}{\rangle}{#1 \delimsize\vert #2}

\hypersetup{citecolor=blue}
\newcommand{\dif}{\mathrm{d}}
\newcommand{\Eqref}[1]{(\ref{#1})}
\newcommand{\half}{\frac{1}{2}}

\newcommand{\brac}[1]{\left(#1 \right)}
\newcommand{\sbrac}[1]{\left[#1\right]}

\begin{document}

\title{Hypocycloid motion in the Melvin magnetic universe}

\author{Yen-Kheng Lim\footnote{Email: yenkheng.lim@gmail.com}}

\affil{\normalsize{\textit{Department of Mathematics, Xiamen University Malaysia, 43900 Sepang, Malaysia}}}

\date{\normalsize{\today}}
\maketitle
 
\renewcommand\Authands{ and }
\begin{abstract}
 The trajectory of a charged test particle in the Melvin magnetic universe is shown to take the form of hypocycloids in two different regimes, the first of which is the class of perturbed circular orbits, and the second of which is in the weak-field approximation. In the latter case we find a simple relation between the charge of the particle and the number of cusps. These two regimes are within a continuously connected family of deformed hypocycloid-like orbits parametrised by the magnetic flux strength of the Melvin spacetime. 
\end{abstract}


\section{Introduction} \label{intro}

The Melvin universe describes a bundle of parallel magnetic field lines held together under its own gravity in equilibrium \cite{Melvin:1963qx,Melvin:1965zza}. The possibility of such a configuration was initially considered by Wheeler \cite{Wheeler:1955}, and a related solution was obtained by Bonnor \cite{Bonnor:1954}, though in today's parlance it is typically referred to as the \emph{Melvin spacetime} \cite{Griffiths:2009dfa}. By the duality of electromagnetic fields, a similar solution consisting of parallel \emph{electric} fields can be obtained. In this paper, we shall mainly be interested in the magnetic version of this solution. 

The Melvin spacetime has been a solution of interest in various contexts of theoretical high-energy physics. For instance, the Melvin spacetime provides a background of a strong magnetic field to induce the quantum pair creation of black holes \cite{Garfinkle:1993xk,Dowker:1993bt}. Havrdov\'{a} and Krtou\v{s} showed that the Melvin universe can be constructed by taking the two charged, accelerating black holes and pushing them infinitely far apart \cite{Havrdova:2006gi}. More recently, the generalisation of the solution to include a cosmological constant has been considered in \cite{Astorino:2012zm,Lim:2018vbq,Zofka:2019yfa}.

Aside from Melvin and Wallingford's initial work \cite{doi:10.1063/1.1704937} and that of Thorne \cite{PhysRev.139.B244}, the motion of test particles in a magnetic universe was typically studied in a more general setting of the Ernst spacetime \cite{Ernst:1975}, which describes a black hole immersed in the Melvin universe. The motion of particles in this spacetime was studied in \cite{Dadhich1979,Estaban1984,Karas1990,Karas1992,Dhurandhar:1983,Stuchlik:2008xi,Lim:2015oha,Li:2018wtz,Nurmagambetov:2018het,Tursunov:2018udx,Pavlovic:2019rim}, and also the magnetised naked singularity was studied in Ref.~\cite{Babar:2015kaa}. 
The study of charged particles in the Ernst spacetime has also informed works in other related areas such as in Refs.~\cite{Nurmagambetov:2018het,Akram:2018lmt,Heydari-Fard:2019pxd}. 

Of particular relevance to this paper is the interaction between the Lorentz force and the gravitational force acting on an electrically charged test mass. As is well known in many textbooks of electromagnetism, a particle moving in a field of mutually perpendicular electric and magnetic fields will experience a trajectory in the shape of a cycloid \cite{JacksonEM}. In this paper, we focus on a similar situation, except that the electric field will be replaced with a gravitational field. The cycloid-like, or, more generally, trochoid-like motion was obtained by Frolov et al. \cite{Frolov:2010mi,Frolov:2014zia} in the study of charged particles in a weakly-magnetised Schwarzschild spacetime \cite{PhysRevD.10.1680}. A similar motion was considered in the Melvin spacetime by the present author in Ref.~\cite{Lim:2015oha}. In this paper, we will extend this idea further to show that the trajectories are more generally deformed \emph{hypocycloids}, which are curves formed by the locus of a point attached to the rim of a circle that is rolling inside another larger circle.  

Hypocycloid trajectories are well known as solutions to various brachistochrone problems in mechanics. For instance, the path of least time in the interior of a uniform gravitating sphere \cite{Venezian} is a hypocycloid. We will see how the trajectories in Melvin spacetime take a hypocycloidal shape as well, specifically in two different regimes of motion. The first of these is the case of perturbed circular motion, and the second is in the weak-field regime. By considering numerical solutions we see that a generic motion in the non-perturbative case consists of a family of deformed hypocycloids.

While the study of charged particles in strong gravitational and magnetic fields are typically candidates of astrophysical interest, the highly ordered motion with finely tuned parameters considered in this paper is perhaps more of a mathematical interest instead. To this end, it may be interesting to study the mathematical connections between hypocycloids and the equations of motion of the Melvin spacetime. As particle motion is typically studied to reveal the underlying geometry of a spacetime, the fact that the motion here is hypocycloids may yet hint at something about the geometry of the Melvin universe. 

The rest of this paper is organised as follows. In Sec.~\ref{sec_eom}, we review the essential features of the Melvin spacetime and derive the equations of motion for an electrically charged test mass. Subsequently, in Sec.~\ref{sec_trochoid}, we consider the perturbation of circular orbits. This was already briefly studied by the present author in a short subsection in \cite{Lim:2015oha}. Here we will review the earlier results and provide some additional details.  In Sec.~\ref{sec_hypocycloid}, we show that trajectories in the weak-field regime correspond precisely to hypocycloids, as well as the study of numerical solutions beyond the weak-field regime. A brief discussion and closing remarks are given in Sec.~\ref{sec_conclusion}. In Appendix \ref{sec_app}, we review the basic properties of hypocycloids.


\section{Equations of motion} \label{sec_eom}

The Melvin magnetic universe is described by the metric
\begin{align}
 \dif s^2&=\Lambda^2\brac{-\dif t^2+\dif r^2+\dif z^2}+\frac{r^2}{\Lambda^2}\dif\phi^2,\quad \Lambda=1+\frac{1}{4}B^2r^2,
\end{align}
where the magnetic flux strength is parametrised by $B$. The gauge potential giving rise to the magnetic field is 
\begin{align}
 A=\frac{Br^2}{2\Lambda}\dif\phi.
\end{align}
The spacetime is invariant under the transformation
\begin{align}
 B\rightarrow-B,\quad\phi\rightarrow-\phi. \label{symmetry}
\end{align}
Therefore we can consider $B\geq0$ without loss of generality.

We shall describe the motion of a test particle carrying an electric charge $e$ by a parametrised curve $x^\mu(\tau)$, where $\tau$ is an appropriately chosen affine parameter. In this paper, we will mainly be considering time-like trajectories, for which $\tau$ can be taken to be the particle's proper time. The trajectory is governed by the Lagrangian $\mathcal{L}=\half g_{\mu\nu}\dot{x}^\mu\dot{x}^\nu+eA_\mu\dot{x}^\mu$, where over-dots denote derivatives with respect to $\tau$. In the Melvin spacetime, the Lagrangian is explicitly
\begin{align}
 \mathcal{L}&=\half\sbrac{\Lambda^2\brac{-\dot{t}^2+\dot{r}^2+\dot{z}^2}+\frac{r^2}{\Lambda^2}\dot{\phi}^2}+\frac{eBr^2}{2\Lambda}\dot{\phi}.
\end{align}
Since $\partial_t$, $\partial_z$, and $\partial_\phi$ are Killing vectors of the spacetime, we have the first integrals
\begin{align}
 \dot{t}=\frac{E}{\Lambda^2},\quad\dot{z}=\frac{P}{\Lambda^2},\quad\dot{\phi}=\frac{\Lambda^2}{r^2}\brac{L-\frac{eBr^2}{2\Lambda}}, \label{tdotzdotphidot}
\end{align}
where $E$, $P$, and $L$ are constants of motion which we shall refer to as the particle's energy, linear momentum in the $z$ direction, and angular momentum, respectively. 

To obtain an equation of motion for $r$, we use the invariance of the inner product of the 4-velocity $g_{\mu\nu}\dot{x}^\mu\dot{x}^\nu=\epsilon$. For time-like trajectories, one can appropriately rescale the affine parameter such that $\epsilon=-1$. Inserting the components of the metric, this gives 
\begin{align}
 \Lambda^4\dot{r}^2=E^2-P^2-V^2_{\mathrm{eff}}, \label{Veff}
\end{align}
where $V_{\mathrm{eff}}^2$ is the effective potential
\begin{align}
 V_{\mathrm{eff}}^2=\frac{\Lambda^4}{r^2}\brac{L-\frac{eBr^2}{2\Lambda}}^2+\Lambda^2.
\end{align}
Another equation of motion for $r$ can be obtained by applying the Euler--Lagrange equation $\frac{\dif}{\dif\tau}\frac{\partial\mathcal{L}}{\partial\dot{r}}=\frac{\partial\mathcal{L}}{\partial r}$, which leads to a second-order differential equation 
\begin{align}
 \ddot{r}&=-\frac{\Lambda'}{\Lambda}\dot{r}^2+\brac{P^2-E^2}\frac{\Lambda'}{\Lambda^5}+\frac{1}{r^3}\brac{1-\frac{r\Lambda'}{\Lambda}}\brac{L-\frac{eBr^2}{2\Lambda}}^2\nonumber\\
     &\quad\hspace{2cm}+\frac{eB\Lambda}{r}\brac{1-\frac{r\Lambda'}{2\Lambda}}\brac{L-\frac{eBr^2}{2\Lambda}},\label{rddot}
\end{align}
where primes denote derivatives with respect to $r$. Another useful equation can be obtained by taking $\frac{\dif r}{\dif\phi}=\frac{\dot{r}}{\dot{\phi}}$, which gives 
\begin{align}
 \brac{\frac{\dif r}{\dif\phi}}^2&=\frac{r^4\brac{E^2-P^2-V_{\mathrm{eff}}^2}}{\Lambda^8\brac{1-\frac{eBr^2}{2\Lambda}}^2}. \label{drdphi}
\end{align}
To obtain the trajectory of the particle, one can solve either Eq.~\Eqref{rddot} or \Eqref{Veff} to obtain $r$. Along with the integrations of Eq.~\Eqref{tdotzdotphidot}, one completely determines the particle motion. 

We note that the metric is invariant under Lorentz boosts along the $z$ direction. Therefore, we can always choose a coordinate frame in which the particle is located at $z=\mathrm{constant}$. This is equivalent to fixing $P=0$ without loss of generality. Furthermore, the equation for $\dot{\phi}$ in Eq.~\Eqref{tdotzdotphidot} is invariant under the sign change $L\rightarrow-L$ if the transformation is accompanied by Eq.~\Eqref{symmetry}. Therefore we shall consider $L\geq0$ without loss of generality as well.

For an appropriately chosen range of $E$ and $L$, the allowed range of $r$ can be specified by the condition that $\dot{r}^2\geq 0$, or, equivalently, $E^2-V_{\mathrm{eff}}^2\geq 0$. We denote this range by 
\begin{align}
 r_-\leq r\leq r_+, \label{allowed_r}
\end{align}
where $r_\pm$ are two positive real roots of the equation $E^2-V_{\mathrm{eff}}^2=0$. For given values of $B$, $e$, and $L$, the minima of $V_{\mathrm{eff}}^2$ gives the circular orbit $r=r_0$, which is the root of $\frac{\dif (V_{\mathrm{eff}}^2)}{\dif r}=0$, where 
\begin{align}
 \frac{\dif (V_{\mathrm{eff}}^2)}{\dif r}&=\frac{4+B^2r^2}{128r^3}\big[(3B^2r^2-4)(4+B^2r^2)L^2\nonumber\\
 &\hspace{3cm} -12B^3r^4e(4+B^2r^2)L+4B^2r^4(8+3r^2B^2e^2+4e^2)\big]. \label{dVdr}
\end{align}

An important quantity for the context of this paper is the value of $r$ where $\dot{\phi}$ vanishes. Denoting this value as $r_*$, we have, using Eq.~\Eqref{tdotzdotphidot},  
\begin{align}
 L=\frac{eBr_*^2}{2\Lambda_*}\quad\leftrightarrow\quad r_*=2\sqrt{\frac{L}{B(2e-LB)}}, \label{L_eqn}
\end{align}
where we have denoted $\Lambda_*=1+\frac{1}{4}B^2r_*^2$. Given $r_*$, one can determine $L$ from above, or vice versa. We note that \Eqref{L_eqn} requires $e$ and $L$ to carry the same sign. Since we have used the symmetry of the spacetime to fix $L\geq0$, the existence of $r_*$ then requires $e\geq0$ as well. Substituting Eq.~\Eqref{L_eqn} into \Eqref{dVdr}, we find 
\begin{align*}
 \left.\frac{\dif(V_{\mathrm{eff}}^2)}{\dif r}\right|_{r=r_*}=\frac{4eB\sqrt{L}}{2e-LB}=\frac{eB^2r_*^2}{\sqrt{L}}.
\end{align*}
As $e$ and $L$ are both positive, the above equation shows that $r=r_*$ must lie within the range where $V_{\mathrm{eff}}^2$ has a positive slope, which is $r_*\geq r_0$. Another quantity of interest is the value of $V_{\mathrm{eff}}^2$ at $r=r_*$. We shall denote this as $E_*^2=\left.V_{\mathrm{eff}}^2\right|_{r=r_*}$

\begin{figure}
 \begin{center}
 \begin{subfigure}[b]{0.49\textwidth}
   \begin{center}
   \includegraphics{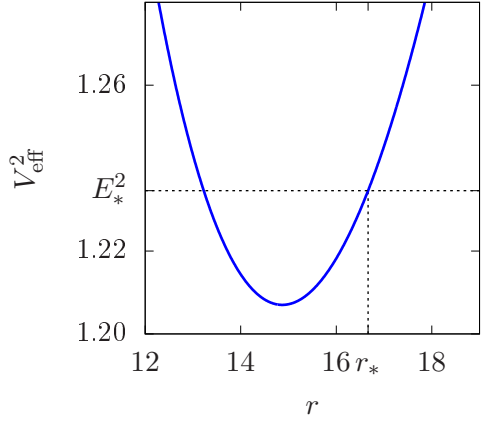}
   \caption{$V_{\mathrm{eff}}^2$ vs $r$.}
   \label{fig_rep_Veff}
   \end{center}
  \end{subfigure}
  \begin{subfigure}[b]{0.49\textwidth}
  \begin{center}
   \includegraphics{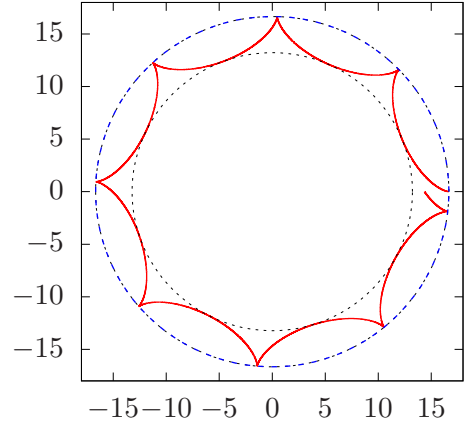}
   \caption{$E^2=E_*$. (Common.)}
   \label{fig_rep_common}
   \end{center}
  \end{subfigure}
  \begin{subfigure}[b]{0.49\textwidth}
  \begin{center}
   \includegraphics{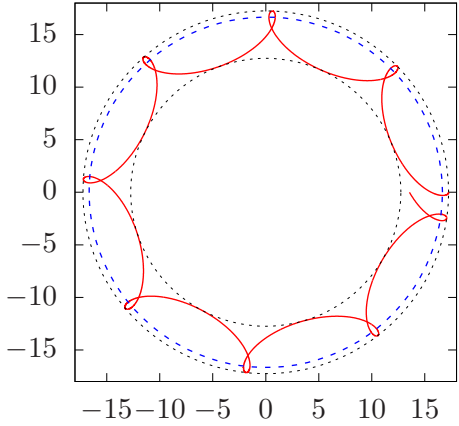}
   \caption{$E^2=E_*+0.02$. (Prolate.)}
   \label{fig_rep_prolate}
   \end{center}
  \end{subfigure}
  \begin{subfigure}[b]{0.49\textwidth}
  \begin{center}
   \includegraphics{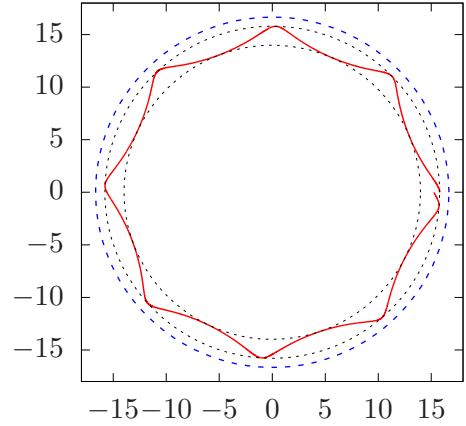}
   \caption{$E^2=E_*-0.02$. (Curtate.)}
   \label{fig_rep_curtate}
   \end{center}
  \end{subfigure}
  \caption{(Colour online.) Examples of orbits with hypocycloid-like behaviour, with $B=0.04$, $e=2$, and $L=10$. For this value of angular momentum, $r_*=16.66667$, and $E_*^2=V_{\mathrm{eff}}^2(r_*)=1.2346$. Figure \ref{fig_rep_Veff} shows the effective potential as a function of $r$, and Fig.~\ref{fig_rep_common}, \ref{fig_rep_prolate}, and \ref{fig_rep_curtate} show orbits with energies $E=E_*$, $E>E_*$, and $E<E_*$, plotted in Cartesian-like coordinates $X=r\cos\phi$, $Y=r\sin\phi$. The two black dotted circles are $r=\pm$, the boundaries of the ranges of allowed $r$ where $\dot{r}^2\geq0$. The blue dashed circles are $r=r_*$.}
  \label{fig_rep}
\end{center}

\end{figure}

We now briefly explain the significance of the quantity $r_*$, using a representative example of $B=0.04$, $e=2$, and $L=10$ shown in Fig.~\ref{fig_rep}. For $L=10$, we use Eq.~\Eqref{L_eqn} to obtain $r_*\simeq1.6667$.\footnote{We shall use the symbol $\simeq$ to indicate that the displayed numerical values are precise up to five decimal places.}  Now, for different choices of $E$, the resulting range \Eqref{allowed_r} may or may not contain $r_*$. There are three possible cases.

In the first case, we have $r_*=r_+$. This occurs when the particle carries an energy $E=E_*$. In this case, $\dot{\phi}$ vanishes the moment it reaches maximum radius where $\dot{r}=0$. The orbit forms a sharp cusp at $r=r_+$, as the one shown in Fig.~\ref{fig_rep_common}. In the case $r_-<r_*<r_+$, the derivative $\dot{\phi}$ will change sign upon crossing $r=r_*$, and change again on its return crossing. This results in the orbit curling up into a coil-like structure, shown in Fig.~\ref{fig_rep_prolate}. Finally, for $r_*>r_+$, the point $r_*$ is not accessible by the particle. Therefore $\dot{\phi}$ does not vanish. Rather, it oscillates between finite, non-zero values. The resulting orbits have a sinusoidal appearance such as in Fig.~\ref{fig_rep_curtate}.

\section{Perturbations of circular orbits} \label{sec_trochoid}

The equations of motion can be solved by $r=\mathrm{constant}=r_0$, corresponding to circular orbits. In order to satisfy \Eqref{rddot} and \Eqref{Veff}, the energy and angular momentum are required to be 
\begin{align}
 E^2&=\frac{(4L+2eBr_0^2-LB^2r_0^2)(4L-2eBr_0^2+LB^2r_0^2)(4+B^2r_0^2)^3}{512B^2r_0^4},\label{E0}\\
 L&=\frac{2\brac{3B^2er_0^2\pm 2\sqrt{4e^2+8-6B^2r_0^2}}r^2_0B}{(3B^2r_0^2-4)(4+B^2r_0^2)}. \label{L0}
\end{align}
Equivalently, Eq.~\Eqref{L0} can be obtained by solving \Eqref{dVdr} for $L$, then substituting the results along with $\dot{r}=0$ into \Eqref{Veff} to obtain $E^2$. 
In the following we shall take the lower sign for \Eqref{L0}, as this is the case that will be related to hypocycloid motion of interest in this paper.

Next, we perturb about the circular orbits by writing $r$ in the form 
\begin{align}
 r(\tau)&=r_0+\varepsilon r_1(\tau). \label{r_perturb}
\end{align}
Further expressing $E$ and $L$ in terms of $e$, $B$ and $r_0$ via \Eqref{E0} and \Eqref{L0}, expanding Eq.~\Eqref{rddot} in $\varepsilon$, we find that the first-order terms describe a harmonic oscillator,
\begin{align}
 \ddot{r}_1&=-\omega^2r_1,
\end{align}
where 
\begin{align}
 \omega^2=\frac{2\brac{3B^6r_0^6L^2-12eB^5r_0^6L+12e^2B^4r_0^6-16B^2r_0^2L^2+128L^2}}{r_0^4\brac{4+B^2r_0^2}^3}.
\end{align}
Subsequently, we expand \Eqref{tdotzdotphidot} to obtain
\begin{align}
 \dot{\phi}&=\beta_0-\beta_1\varepsilon r_1+\mathcal{O}\brac{\varepsilon^2}, \label{phi_perturb}
\end{align}
where 
\begin{subequations}
\begin{align}
 \beta_0&=\frac{(4L+LB^2r_0^2-2eBr_0^2)(4+B^2r_0^2)}{16r_0^2},\\
 \beta_1&=\frac{16L-LB^6r_0^4+2eB^3r_0^4}{8r_0^3}.
\end{align}
\end{subequations}
In particular, $\beta_0$ is the angular frequency of revolution of the unperturbed circular motion (the cyclotron frequency). With this, we find that the solution to Eqs.~\Eqref{r_perturb} and \Eqref{phi_perturb} are
\begin{subequations}
\begin{align}
 r&=r_0+\varepsilon\cos\omega\tau+\mathcal{O}\brac{\varepsilon^2},\label{trochoid_r}\\
 \phi&=\frac{\beta_0}{\omega}\brac{\omega\tau-\zeta\sin\omega\tau}+\mathcal{O}\brac{\varepsilon^2}, \label{trochoid_phi}
\end{align}
\end{subequations}
where 
\begin{align}
 \zeta=\frac{\beta_1\varepsilon}{\beta_0}. \label{eta_pert}
\end{align}
Neglecting the terms second order in $\varepsilon$ and beyond, we have the equation of a family of trochoids parametrised by $\zeta$. Recalling the standard description of trochoids, the case $\zeta>1$, describes the \emph{prolate cycloid}, $\zeta<1$ describes the \emph{curtate cycloid}, and $\zeta=1$ corresponds to the \emph{common cycloid}. 
\begin{figure}
 \begin{center}
 \begin{subfigure}[b]{0.9\textwidth}
   \begin{center}
   \includegraphics{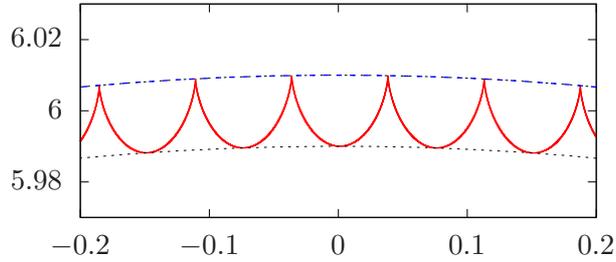}
   \caption{$e=18.068$, $\eta\simeq1.0903\approx 1$. (Common cycloid.)}
   \label{fig_trochoid_common}
   \end{center}
  \end{subfigure}
  \begin{subfigure}[b]{0.9\textwidth}
  \begin{center}
   \includegraphics{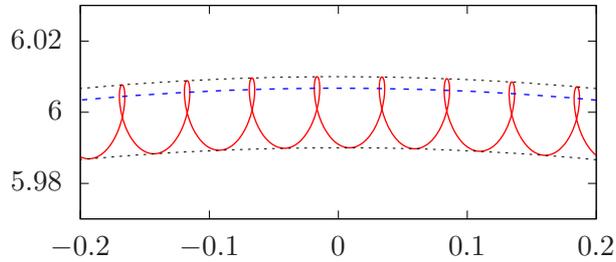}
   \caption{$e=22$, $\eta\simeq1.4840>1$. (Prolate cycloid.)}
   \label{fig_trochoid_prolate}
   \end{center}
  \end{subfigure}
  \begin{subfigure}[b]{0.9\textwidth}
  \begin{center}
   \includegraphics{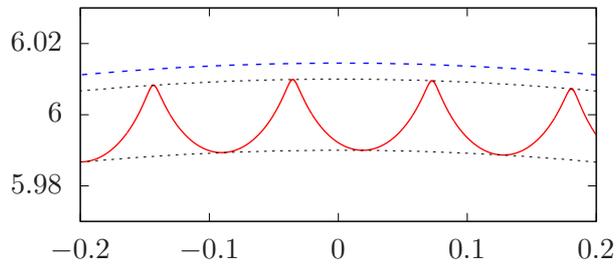}
   \caption{$e=15$, $\eta\simeq0.69100<1$. (Curtate cycloid)}
   \label{fig_trochoid_curtate}
   \end{center}
  \end{subfigure}
  \caption{(Colour online.) Sections of perturbed circular orbits about $r_0=6$ for $B=0.1$ and various $e$. Here, we take $\lambda=0.01$ and the angular momentum of these orbits are exactly equal to its unperturbed case calculated with \Eqref{L0}, whereas the energies of the orbits are obtained by solving \Eqref{Veff} for $E$. As in Fig.~\ref{fig_rep}, the black dotted arcs denote the boundaries of the allowed range $r_-\leq r\leq r_+$ and the blue dashed arc indicates the point $r=r_*$ where $\dot{\phi}$ vanishes.}
  \label{fig_trochoid}
\end{center}

\end{figure}

Recall that the standard cycloid is formed by the locus of a point on a circle rolling on a flat plane.  In the present case, this `plane' is not flat but rather a large circle of radius $\sim r_0$, and it only approximates a flat plane for intervals of motion of order $\varepsilon\ll r_0$. Figure\ref{fig_trochoid} shows the zoomed-in sections of perturbed circular orbits about $r_0=6$ for a spacetime of magnetic flux parameter $B=0.1$ and various values of $e$. The circular orbits are perturbed by $\lambda=0.01$ around $r_0=6$. We see that, depending on the charge of the particle, the perturbed orbit can either be a common cycloid ($e=18.068$, Fig.~\ref{fig_trochoid_common}), prolate cycloid ($e>18.068$, Fig.~\ref{fig_trochoid_prolate}), or curtate cycloid ($e<18.068$, Fig.~\ref{fig_trochoid_curtate}).

As $\phi$ evolves across a period of $2\pi$, the number of $r$ oscillations is approximately
\begin{align}
 n=\frac{\omega}{\beta_0}. \label{n_pert}
\end{align}
We can calculate $n$ for the examples shown in Fig.~\ref{fig_trochoid}. For the parameters giving the common cycloid in Fig.~\ref{fig_trochoid_common}, we have $n\simeq 506.1$. For prolate cycloid of Fig.~\ref{fig_trochoid_prolate} it is $n\simeq 749.4$. Finally, for the curtate cycloid in Fig.~\ref{fig_trochoid_curtate}, $n\simeq349.4$. In the regime of perturbed circular orbits, the quantity $n$ defined in \Eqref{n_pert} is the number of cusps formed as $\phi$ goes through one period of $2\pi$.

Of course, the locus of a point on a circle rolling inside a larger circle is also well-known curve called the \emph{hypotrochoid}. For the rest of the paper we shall focus on the case of the \emph{common hypocycloid}, which is the analogue to the common cycloid and is also characterised by the occurence of sharp cusps. In the next section, we will show how the hypocycloid can be extracted from the equations of motion beyond the regime of perturbed circular orbits.

\section{Hypocycloid-like trajectories} \label{sec_hypocycloid}

Like their analogues in cycloids, the hypocycloids are characterised by their trajectories having sharp cusps at maximum radius. In terms of the equations of motion, this corresponds to $\dot{r}$ and $\dot{\phi}$ being zero simultaneously. In other words,
\begin{align}
 r_+=r_*. \label{common_r*}
\end{align} 
In this case, the required value of $E$ for the orbit to be a common hypocycloid is obtained by substituting $r=r_+=r_*$ into Eq.~\Eqref{Veff}. At this position, the radial velocity is zero. Therefore we put $\dot{r}=0$ and solve for $E$ to obtain 
\begin{align}
 E=\Lambda_*. \label{common_E}
\end{align}
Having the energy and angular momentum fixed by $r_*$, the equations of motion now become 
\begin{align}
 \brac{\frac{\dif r}{\dif\phi}}^2&=\frac{4r^4}{e^2B^2\Lambda^4\brac{\frac{r_*^2}{\Lambda_*}-\frac{r^2}{\Lambda}}}\sbrac{\frac{\Lambda_*^2}{\Lambda^4}-\frac{1}{\Lambda^2}-\frac{eB}{2r^2}\brac{\frac{r_*^2}{\Lambda_*}-\frac{r^2}{\Lambda}}}. \label{common_drdphi}
\end{align}
When the magnetic field is weak, we will now show that the trajectory can be approximated by hypocycloids. To this end, we take $B$ to be small while keeping $e$ sufficiently large so that the gravitational effects of the magnetic field is reduced while keeping the Lorentz interaction on the charged particle significant. Therefore, we introduce the parametrisation
\begin{align}
 B=g\lambda^{2},\quad e=\frac{q}{\lambda},
\end{align}
for some constants $g$ and $q$, and expand in small $\lambda$. Then, Eq.~\Eqref{Veff} becomes 
\begin{align}
 \dot{r}^2&=g^2\lambda^2\sbrac{-\frac{1}{4}\brac{q^2+2\lambda^2}r^2+\half\brac{q^2+\lambda^2} r_+^2-\frac{q^2r_+^2}{r^2}}+\mathcal{O}\brac{\lambda^6g^4q^2}\nonumber\\
   &=\frac{1}{4}g^2\lambda^2\brac{q^2+2\lambda^2}\frac{1}{r^2}\brac{r_+^2-r^2}\brac{r^2-\frac{r_+^2q^2}{q^2+2\lambda^2}}+\mathcal{O}\brac{\lambda^6g^4q^2}, \label{common_rdot_exp}
\end{align}
while the equation of motion for $\phi$ gives 
\begin{align}
 \dot{\phi}&=\half gq\lambda\frac{r_+^2-r^2}{r^2}+\mathcal{O}\brac{g^2q^2\lambda^6}, \label{common_phidot_exp}
\end{align}
and Eq.~\Eqref{common_drdphi} is similarly expanded in small $\lambda$ to become
\begin{align}
 \brac{\frac{\dif r}{\dif\phi}}^2=-r^2+\frac{2r^4}{r_+^2-r^2}\frac{\lambda^2}{q^2}+\mathcal{O}\brac{g^2\lambda^4}. \label{common_drdphi_exp}
\end{align}
Equivalently, one could also obtain Eq.~\Eqref{common_drdphi_exp} by dividing $\dot{r}^2/\dot{\phi}^2$ using the expressions from \Eqref{common_rdot_exp} and \Eqref{common_phidot_exp} while neglecting the higher-order terms and rearranging. 

Ignoring the higher-order terms, Eqs.~\Eqref{common_rdot_exp}, \Eqref{common_phidot_exp}, and \Eqref{common_drdphi_exp} are precisely the standard equations of the hypocycloid given in Eqs.~\Eqref{commonhyp_rdot}, \Eqref{commonhyp_phidot}, and \Eqref{commonhyp_drdphi}, upon the identifying the parameters as 
\begin{align}
 \frac{2\lambda^2}{q^2}&=\frac{r_+^2-r_-^2}{r_-^2}=\frac{4(n-1)}{(n-2)^2},\label{lambdaq_identify}\\
 gq\lambda&=\frac{2r_-}{r_+-r_-}.
\end{align}
In the notation of Appendix \ref{sec_app}, we recall that the hypocycloid is a curve traced out by a point sitting on a circle of radius $b$ rolling inside a larger circle of radius $a=bn$. If $n$ is an integer with $n\geq 2$, we get a periodic hypocycloid with $n$ cusps. In terms of $r_\pm$, we have $r_-=r_+-2b$ and $r_+=bn$. Furthermore, as $e=\frac{q}{\lambda}$, Eq.~\Eqref{lambdaq_identify} leads to 
\begin{align}
 e&=\frac{n-2}{\sqrt{2(n-1)}}. \label{common_e}
\end{align}
In the regime of small $\lambda$, this gives the the required charge $e$ for the particle to execute a periodic hypocycloid with $n$ cusps. 

\begin{figure}
 \begin{center}
 \begin{subfigure}[b]{0.49\textwidth}
   \begin{center}
   \includegraphics{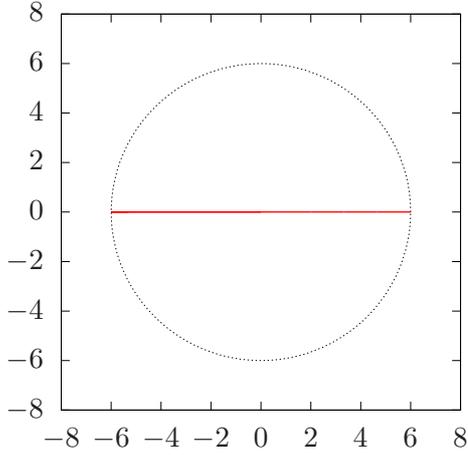}
   \caption{$n=2$, $e=0$ (straight line).}
   \label{fig_common-n2}
   \end{center}
  \end{subfigure}
  \begin{subfigure}[b]{0.49\textwidth}
  \begin{center}
   \includegraphics{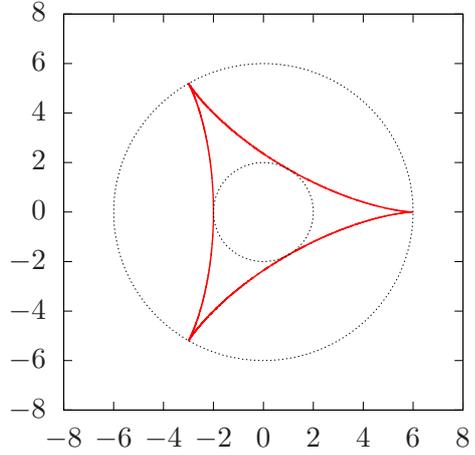}
   \caption{$n=3$, $e=\frac{1}{2}$ (deltoid).}
   \label{fig_common-n3}
   \end{center}
  \end{subfigure}
  \begin{subfigure}[b]{0.49\textwidth}
  \begin{center}
   \includegraphics{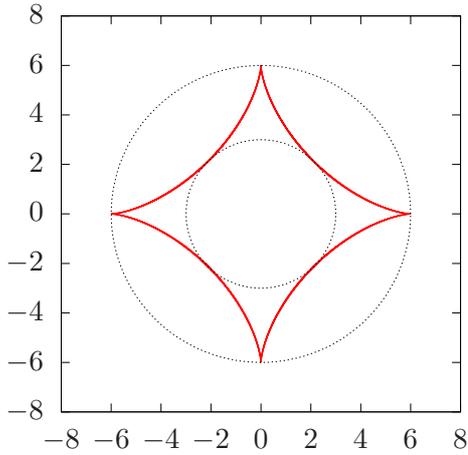}
   \caption{$n=4$, $e=\frac{\sqrt{6}}{3}$ (astroid).}
   \label{fig_common-n4}
   \end{center}
  \end{subfigure}
  \begin{subfigure}[b]{0.49\textwidth}
  \begin{center}
   \includegraphics{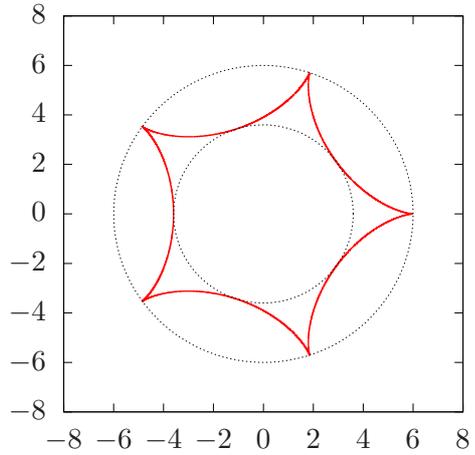}
   \caption{$n=5$, $e=\frac{3\sqrt{2}}{4}$.}
   \label{fig_common-n5}
   \end{center}
  \end{subfigure}
  \caption{(Colour online.) Hypocycloid trajectories for $n=2,\ldots,5$, and their corresponding charges $e$ determined from Eq.~\Eqref{common_e}. The parameters used are $B=0.001$, $r_+=r_*=6$. The energy and angular momentum of each orbit are obtained from Eqs.~\Eqref{common_E} and \Eqref{L_eqn}, respectively. The dotted circles are the boundaries of the range for $r_-\leq r\leq r_+$ of each trajectory.}
  \label{fig_common}
\end{center}

\end{figure}

A straight line segment is technically  a `hypocycloid' with $n=2$. By Eq.~\Eqref{common_e}, this will be the trajectory of a neutral particle with zero angular momentum undergoing radial oscillations about axis of symmetry in the Melvin universe, such as in Fig.~\ref{fig_common-n2}. For $n=3$, Eq.~\Eqref{common_e} tells us that a particle with charge $e=\half$ traces the shape of a hypocycloid with three cusps, called a \emph{deltoid}. (See Fig.~\ref{fig_common-n3}.) For $n=4$, we have a particle with charge $e=\frac{\sqrt{6}}{3}$ tracing out an \emph{astroid}, which is a hypocycloid with four cusps. (See Fig.~\ref{fig_common-n4}.) This follows higher $n$.

To summarise, one can obtain hypocycloid trajectories as follows. Given a choice of $r_*=r_+$ and $n$, the requisite energy and angular momentum are calculated from Eqs.~\Eqref{common_E} and \Eqref{L_eqn}. The charge of the particle is fixed by Eq.~\Eqref{common_e}. One also has to choose the magnetic field strength $B$ so that terms of order $\mathcal{O}\brac{e^2B^2}$ are sufficiently small. In this way, the higher-order terms of Eqs.~\Eqref{common_rdot_exp}, \Eqref{common_phidot_exp} and \Eqref{common_drdphi_exp} can be neglected. This ensures that the equations of motion hold up to reasonable precision as hypocycloid equations. 

We can verify the above arguments by solving the full non-perturbative equations of motion numerically. In other words, for a choice of $r_*$, $n$, and a small $B$, we integrate Eqs.~\Eqref{rddot} and \Eqref{tdotzdotphidot} using a fourth-order Runge--Kutta method. In Fig.~\ref{fig_common}, we obtain the trajectories for $B=0.001$ and $r_*=r_+=6$. For these values, the deviation of the trajectory from being a true exact hypocycloid is of $\mathcal{O}\brac{e^2B^2}\sim\mathcal{O}\brac{10^{-5}}$. With this relatively small error, the visual appearance of the orbits in Fig.~\ref{fig_common} indeed resembles the standard hypocycloid.

Next, we shall explore the shape of the orbits as we increase $B$ to beyond the weak-field regime. As $B$ is increased, the higher-order corrections in Eqs.~\Eqref{common_rdot_exp}, \Eqref{common_phidot_exp}, and \Eqref{common_drdphi_exp} become important, and will no longer coincide with the hypocycloid equations. Nevertheless, we are still able to solve the non-perturbative equations numerically and explore its behaviour. 

\begin{figure}
 \begin{center}
 \begin{subfigure}[b]{0.49\textwidth}
   \begin{center}
   \includegraphics{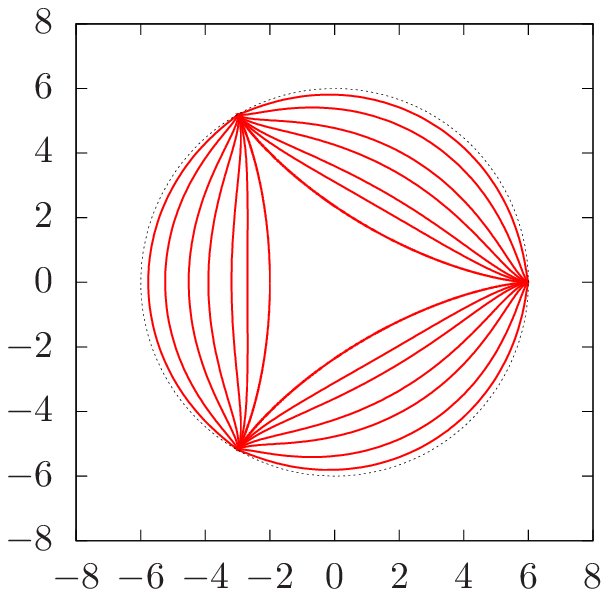}
   \caption{$n=3$.}
   \label{fig_deform-n3}
   \end{center}
  \end{subfigure}
  \begin{subfigure}[b]{0.49\textwidth}
  \begin{center}
   \includegraphics{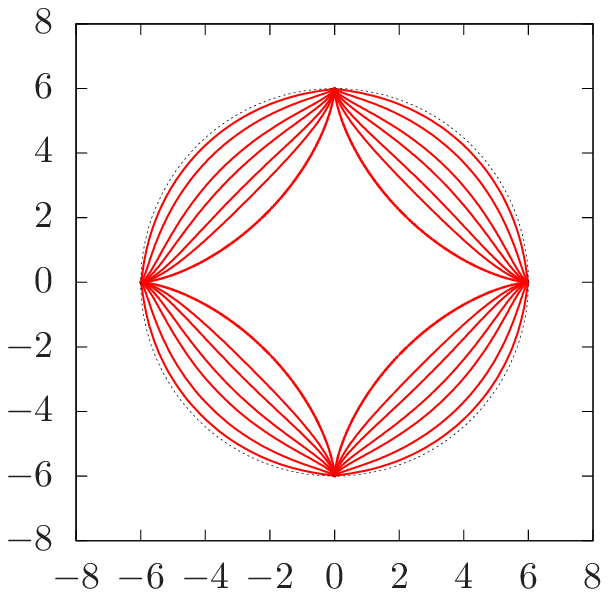}
   \caption{$n=4$.}
   \label{fig_deform-n4}
   \end{center}
  \end{subfigure}
  \caption{(Colour online.) Sequence of hypocycloid-like orbits of increasing $B$. Starting from the innermost orbits to the outermost one, the corresponding values of $B$ are $0.001$, $0.16$, $0.20$, $0.25$, $0.30$, $0.40$, and $0.60$.}
  \label{fig_deform}
\end{center}

\end{figure}

As demonstrated in Fig.~\ref{fig_deform}, as $B$ increases, the hypocycloids are continuously deformed. The innermost curve curve is the one that most closely approximates hypocycloids with $B=0.001$ and $e$ given by Eq.~\Eqref{common_e}. The subsequent curves are obtained by increasing $B$ and tuning $e$ manually until we obtain the periodic orbit with desired number of cusps. We see that as $B$ increases, the segments of curves joining two cusps are deformed from a concave shape into a convex one. Furthermore, the range of allowed radii $r_-\leq r\leq r_+$ becomes narrower as $B$ increases, until we see that the outermost orbit with the largest $B$, the orbits begin to resemble a circular orbit. In the case of Fig.~\ref{fig_deform}, the outermost orbit depicted is for $B=0.6$.

In fact, we can check that these orbits of large $B$ correspond to the perturbed circular orbits of Sec.~\ref{sec_trochoid}. We do so by checking that the numerical (non-perturbative) solution matches the perturbed solution of Sec.~\ref{sec_trochoid}. For instance, let us take the $n=3$ case shown in Fig.~\ref{fig_deform-n3}. The outermost $n=3$ orbit at $B=0.6$ is formed by a particle carrying charge $e\simeq10.6204$. The maximum and minimum radii of the motion are $r_+=6.0=r_*$ and $r_-\simeq5.7674$, respectively. We shall treat this as a reasonably narrow range such that the orbit is regarded as a perturbed circular orbit about $r_0\approx\frac{r_++r_-}{2}\simeq5.8837$, and the perturbation parameter is $\varepsilon\approx\frac{r_+-r_-}{2}\simeq0.1163$. Inserting these values into Eqs.~\Eqref{n_pert} and \Eqref{eta_pert}, we obtain 
\begin{align}
 n\simeq3.0057,\quad \zeta\simeq 1.0214,
\end{align}
which is consistent with a common cycloid ($\zeta=1$) performing three oscillations in $r$ within one angular period, thus forming $n=3$ cusps.

Performing a further check for the $n=4$ orbit of Fig.~\ref{fig_deform-n4}, we have $B=0.6$, $e\simeq 12.2849$. The maximum and minimum radii are $r_+=6.0=r_*$ and $r_-=5.8275$, for which we take $r_0\simeq 5.9138$ and $\varepsilon\simeq 0.0862$ Inserting these into Eq.~\Eqref{n_pert} and \Eqref{eta_pert}, we find
\begin{align}
 n\simeq 4.0041,\quad \eta\simeq 1.0154,
\end{align}
which is consistent with a common cycloid performing four oscillations in $r$ within one angular period, resulting in $n=4$ cusps.

Similar checks can be performed for higher $n$. Hence, we conclude that the periodic orbits with sharp cusps ($r_+=r_*$) form a family of deformed hypocycloidal curves. One end of this family consists of hypocycloids in the weak-field regime, and on the other end are common cycloids as the perturbation of circular orbits.

\section{Conclusion} \label{sec_conclusion}

In this paper, we have studied a particular type of motion performed by a charged particle in the Melvin spacetime. It was found that in two different regimes, the trajectory takes the shape of a hypocycloid. The first regime where this occurs is in the class of perturbed circular orbits \cite{Lim:2015oha}, and the second is in the weak field approximation. Particularly, in the latter case, we find that the particle's charge $e$ is related to the number of cusps $n$ of the hypocycloid by Eq.~\Eqref{common_e}. 

The trajectories in the two regimes are continuously connected by a family of deformed trajectories that still retain the features of the hypocycloid, namely its configuration of cusps. 
This family of intermediate solutions are obtained non-perturbatively via numerical solutions. We have seen that as $B$ increases beyond the weak field regime, the hypocycloids are deformed until it arrives at the regime of hypocycloids in the perturbed circular orbit regime.

As the hypocycloid equations were extracted from two different perturbations of the equations of motion in the Melvin spacetime, one naturally wonders whether there are any more interesting connections to other mathematical properties of the hypocycloids. To briefly speculate along this line of thought, it was recently noted that hypocycloids are related to the positions of eigenvalues of $SU(n)$ in the complex plane \cite{Kaiser_2006}. It may be intriguing to wonder whether this carries any implications in the context of charged particle motion in the Melvin spacetime.

\appendix 

\section{Parametric equations of the hypocycloid} \label{sec_app}

Consider a disk of radius $b$ rolling without slipping inside a larger circle of radius $a=bn$, where $n>1$. Let $P$ be a point on the edge of the disk at distance $b$ from its centre. The curve traced out by $P$ as the disk rolls in the larger circle is a \emph{hypocycloid}. In Cartesian coordinates, its parametric equations are 
\begin{align}
 x&=b\sbrac{(n-1)\cos\tau+\cos(n-1)\tau},\quad y=b\sbrac{(n-1)\sin\tau-\sin(n-1)\tau},\quad \tau\in\mathbb{R}.
\end{align}
Let $r_-$ and $r_+$ be its minimum and maximum distance from the origin. In terms of these parameters, 
\begin{align}
 b=\frac{r_+-r_-}{2},\quad n=\frac{2r_+}{r_+-r_-}. \label{nrp}
\end{align}
We convert to polar coordinates with $x=r\cos\phi$ and $y=r\sin\phi$. In terms of $r$ and $\phi$, one can show that
\begin{align}
 \frac{\dif r}{\dif\tau}&=\frac{r_+}{r_+-r_-}\frac{\sqrt{(r_+^2-r^2)(r^2-r_-^2)}}{r},\label{commonhyp_rdot}\\
 \frac{\dif\phi}{\dif\tau}&=\frac{r_-^2}{r_+-r_-}\frac{\sqrt{r_+^2-r^2}}{r^2}. \label{commonhyp_phidot}
\end{align}
Eliminating the parameter $t$, we have
\begin{align}
 \brac{\frac{\dif r}{\dif\phi}}^2=\frac{r_+^2}{r_-^2}\frac{r^2\brac{r^2-r_-^2}}{r_+^2-r^2}&=-r^2+\frac{r_+^2-r_-^2}{r_-^2}\frac{r^4}{r_+^2-r_-^2}\nonumber\\
                  &=-r^2+\frac{4(n-1)}{(n-2)^2}\frac{r^4}{r_+^2-r_-^2}, \label{commonhyp_drdphi}
\end{align}
where the second line follows from using Eq.~\Eqref{nrp} to express $r_-$ in terms of $n$ and $r_+$, which then results in cancellations of factors of $r_+$.

\section*{Acknowledgments}
This work is supported by Xiamen University Malaysia Research Fund (Grant No. \\XMUMRF/2019-C3/IMAT/0007).

\bibliographystyle{melcyl}

\bibliography{melcyl}

\end{document}